\documentclass[10pt]{article}

\usepackage{graphicx} 
\usepackage{cite} 
\usepackage{changepage} 
\usepackage{tabularx} 
\usepackage{multirow} 
\usepackage[usenames,dvipsnames]{color}

\newcolumntype{x}[1]{
>{\centering\hspace{0pt}}p{#1}}%
\newcommand{\tn}{\tabularnewline}

\makeatletter
\renewcommand{\@biblabel}[1]{\quad#1.}
\makeatother

\date{}

\newcommand{\blank}[1]{}

\begin{document}

\begin{flushleft}
{\Large \textbf{Pervasive variation of transcription factor orthologs contributes to regulatory network evolution}}
\\[1\baselineskip]
Shilpa Nadimpalli$^{1,2}$, 
Anton V. Persikov$^{2}$, 
Mona Singh$^{1,2,\ast}$
\\[1\baselineskip]
\bf{1} Department of Computer Science, Princeton University, Princeton, NJ, USA
\\
\bf{2} Lewis-Sigler Institute for Integrative Genomics, Princeton University, Princeton, NJ, USA
\\[1\baselineskip]
$\ast$ E-mail: mona@cs.princeton.edu
\end{flushleft}

\section*{Abstract}

Differences in transcriptional regulatory networks underlie much of the phenotypic variation observed across organisms. Changes to \emph{cis}-regulatory elements are widely believed to be the predominant means by which regulatory networks evolve, yet examples of regulatory network divergence due to transcription factor (TF) variation have also been observed. To systematically ascertain the extent to which TFs contribute to regulatory divergence, we analyzed the evolution of the largest class of metazoan TFs, Cys${\sf{_2}}$-His${\sf{_2}}$ zinc finger (C2H2-ZF) TFs, across 12 \emph{Drosophila} species spanning $\sim$45 million years of evolution.  Remarkably, we uncovered that a significant fraction of all C2H2-ZF 1-to-1 orthologs in flies exhibit variations that can affect their DNA-binding specificities. In addition to loss and recruitment of C2H2-ZF domains, we found diverging DNA-contacting residues in $\sim$47\% of domains shared between \emph{D. melanogaster} and the other fly species. These diverging DNA-contacting residues, found in $\sim$66\% of the \emph{D. melanogaster} C2H2-ZF genes in our analysis and corresponding to $\sim$24\% of all annotated \emph{D. melanogaster} TFs, show evidence of functional constraint: they tend to be conserved across phylogenetic clades and evolve slower than other diverging residues. These same variations were rarely found as polymorphisms within a population of \emph{D. melanogaster} flies, indicating their rapid fixation. The predicted specificities of these dynamic domains gradually change across phylogenetic distances, suggesting stepwise evolutionary trajectories for TF divergence. Further, whereas proteins with conserved C2H2-ZF domains are enriched in developmental functions, those with varying domains exhibit no functional enrichments. Our work suggests that a subset of highly dynamic and largely unstudied TFs are a likely source of regulatory variation in \emph{Drosophila} and other metazoans.

\blank{
\section*{Author Summary}

The phenotypic differences observed between closely related organisms are thought to be due largely to changes in regulatory networks. Changes in transcriptional networks can occur via mutations in \emph{cis} binding sites, for which there are numerous known examples, as well as via binding specificity variation in transcription factors (TFs), a less studied phenomenon that has been observed primarily in multi-gene families. Though large-scale experimental studies ascertaining the extent to which TFs contribute to regulatory network variation across organisms are lacking and would be time-consuming, computational methods can begin to address this challenge. Here, we present the first systematic, large-scale analysis of DNA-binding specificity evolution in TF orthologs by computationally leveraging specific features of Cys$_2$-His$_2$ zinc finger TFs, the largest class of TFs in animals and major components of their regulatory programs. We find that divergence of binding specificity in 1-to-1 orthologous TFs not  only is pervasive, in contrast to previous assumptions, but also is functionally important and occurs in a gradual, evolutionarily viable manner. We conclude that the diversification of TFs has most likely played a major and largely unstudied role in gene regulatory network evolution in metazoans.
}

\section*{Introduction}

Differences in regulatory networks have been proposed to be one of the major determinants of the phenotypic variations observed across organisms \cite{king1975}. There are two ways by which regulatory networks evolve: changes in \emph{cis} or \emph{trans}. The predominant view is that regulatory evolution results mainly from the gain and loss of binding sites in \emph{cis}-regulatory regions because incremental, evolutionarily viable steps can occur \cite{wray2003,prudhomme2007,stern2008,liao2010}. Mutations in transcription factors (TFs), on the other hand, can affect the expression of multiple genes and are thought therefore to be more likely to have detrimental consequences \cite{britten1969,stern2000,carroll2005a,wray2007}. Nevertheless, case studies of specific biological systems have revealed instances of regulatory divergence stemming from TF variation. These variations include gene loss as well as gene duplication where the subsequent paralogs exhibit gain and loss of effector domains, changes in interactions with other regulatory proteins, or novel TF binding potential \cite{vlad2014,wagner2008,hannenhalli2008,thomas2009,baker2011,nakagawa2013}. Specific cases of variations in non-duplicated TFs are also known; an example of 1-to-1 orthologous plant TFs with differing binding specificities was recently discovered \cite{sayou2014}, along with a homeodomain TF in animals where the addition of a functionally important transcriptional repressor domain is found in insect orthologs \cite{galant2002,ronshaugen2002}.  However, a large-scale experimental study ascertaining the extent to which TF variation may contribute to overall regulatory network evolution is still lacking; it would require determining DNA-binding specificities or genomic occupancies for numerous TFs across a diverse set of organisms. Computational methods can begin to address this challenge by leveraging specific features of TFs.
  
TFs come in distinct structural classes based upon their incorporation of various DNA-binding domains. For many of these domains, the amino acids conferring DNA-binding specificity are known. This provides a platform to assess TF variation via comparative sequence analysis. The Cys$_2$-His$_2$ zinc finger (C2H2-ZF) TFs in particular are an excellent system to probe for variation; further, they constitute the largest group of TFs in higher metazoans \cite{vaquerizas2009}, making up nearly half of all annotated TFs in human, and are major participants in regulatory programs.  C2H2-ZF domains have a conserved modular structure with binding specificity conferred largely by four DNA-contacting residues in the domain's alpha-helix \cite{pabo2001}. In addition, a C2H2-ZF domain can specify a wide range of three or four base pair targets, and tandem arrays of these domains bind contiguous DNA sequences, giving C2H2-ZF genes the ability to recognize an incredibly diverse set of motifs \cite{enuameh2013}.  These features of C2H2-ZFs allow us to make binding specificity predictions of reasonably high quality for this TF family \cite{benos2002,kaplan2005,persikov2014,gupta2014}.

Previous evolutionary analyses of C2H2-ZF genes revealed a dichotomy in conservation patterns of this family. Tandemly-duplicated C2H2-ZF paralogs exhibit differences in their C2H2-ZF and effector domain count and can be highly dynamic across short evolutionary distances \cite{nowick2011}. The subset of C2H2-ZF KRAB repressor regulators in particular have undergone recent, rapid expansion and divergence in primates and show evidence of adaptive evolution in their DNA-binding domains in human \cite{thomas2009,nowick2010,hughes2011}.  However, such divergence has been found primarily in extremely recent and often species-specific expansions of C2H2-ZFs \cite{liu2014}. In contrast, examples of single-copy 1-to-1 orthologous C2H2-ZF genes, particularly those involved in developmental gene regulation, have been shown to be highly conserved across large evolutionary distances \cite{nowick2011,knight2001,seetharam2010,seetharam2013}. \emph{Prdm9}, a C2H2-ZF gene that mediates homologous recombination but is not known to be a TF, is a notable exception to this trend, and is highly dynamic between and within species despite being single-copy \cite{oliver2009,myers2010,berg2011,segurel2011,groeneveld2012}. More generally, it is widely believed that 1-to-1 orthologous TFs tend to maintain their DNA-binding specificities whereas paralogous TFs are free to vary \cite{hoekstra2007}.
 
We have developed a framework to study and quantify variation in orthologous 1-to-1 C2H2-ZF TFs across closely related species. Our framework utilizes the well-understood binding interface of C2H2-ZFs to evaluate binding site specificity changes resulting from C2H2-ZF variation. We apply our framework on the 12 sequenced \emph{Drosophila} species (phylogenetic tree in Figure 1A), as they benefit from complete, high-quality coverage \cite{drosophila2007}. Further, as a result of their $\sim$45 million years of evolutionary divergence \cite{flybase}, they exhibit extensive regulatory variation   \cite{gompel2003,jeong2006} and diversity in terms of morphology, physiology and ecology \cite{markow2007}. The flies are an ideal model organism set for our study because they have several hundred C2H2-ZF genes found in well-established orthologous relationships. This is in contrast to primate genomes where large-scale species-specific expansions complicate orthology determination.

To assess change, we consider only C2H2-ZF orthologous pairs that are in 1-to-1 relationships between \emph{D. melanogaster}, which we use as a reference species, and each of the 11 other fly species.  We find evidence of functional modifications to DNA-binding potential in a significant proportion of C2H2-ZF genes in \emph{D. melanogaster} with 1-to-1 orthologs in the other fly species. Furthermore, these changes often result in increasingly diverse predicted DNA-recognition motifs as evolutionary distance from \emph{D. melanogaster} increases, implying that C2H2-ZF DNA-binding specificities may evolve gradually in evolutionarily viable steps. Our findings challenge the assumption that 1-to-1 orthologous TFs are always highly conserved and provide evidence that binding specificity modifications in non-duplicated TFs may play an important role in the regulatory evolution of \emph{Drosophila} and other higher metazoans.

\section*{Results}

  \subsection*{C2H2-ZF Domains and Orthogroup Dataset}
  The initial step of our framework to assess variation in C2H2-ZFs was to assemble groups of orthologs (orthogroups) of C2H2-ZF genes across the 12 fly species (Figure 1A). We identified all C2H2-ZF domains and sequences in these species using Pfam \cite{pfam} and HMMER \cite{hmmer} and determined 1-to-1 orthogroups from existing Flybase \cite{flybase} annotations. We then augmented this set using the UCSC Genome Browser \cite{ucscgb} whole genome fly alignment, resulting in a complete, high-quality dataset of all C2H2-ZF sequences in the \emph{Drosophila} species (Methods M1-M3).

  C2H2-ZF domains are known to primarily work in tandem to specify DNA motifs \cite{iuchi2001} (Figure S1B), and so we include only those C2H2-ZF genes with 2{\tt{+}} C2H2-ZF domains in our analysis; we refer to these genes as poly-ZF. Tandem C2H2-ZF domains that are separated by canonical linkers --- stretches of 5 to 12 amino acids, most often matching the expression {\tt{TGE[K$|$R]P[F$|$Y]X}} (Figure S1C) --- have the strongest structural evidence for DNA binding \cite{pabo2001,enuameh2013}. We refer to all domains that are bordered by at least one canonical linker, as defined above, to be ``canonically linked.'' In \emph{D. melanogaster}, of the 339 genes with at least one C2H2-ZF domain, 282 have multiple C2H2-ZF domains, and 239 of those contain canonically linked domains.
    
  We found from 339 to 377 genes with at least one C2H2-ZF domain in each of the 12 \emph{Drosophila} species, 282 to 311 of which were poly-ZF (Figure 1B, cols. 1-2), in accordance with previous studies' findings \cite{thomas2009,enuameh2013}. We also confirmed the relative lack of species-specific C2H2-ZF expansions in \emph{D. melanogaster}: 268 (95.0\%) poly-ZF genes in \emph{D. melanogaster} had a 1-to-1 ortholog in at least one other fly species, and 132 (49.3\%) of these were in 1-to-1 relationships across all species. These 268 1-to-1 orthologous poly-ZFs constitute 35.6\% of the estimated 753 TFs in \emph{D. melanogaster} \cite{adryan2006}. In each non-\emph{melanogaster} species, 63.7\% to 82.2\% of poly-ZF genes had a 1-to-1 ortholog in \emph{D. melanogaster} (Figure 1B, col. 3). In the non-reference poly-ZF genes with 1-to-1 orthologs in \emph{D. melanogaster}, we identified 1000{\tt{+}} C2H2-ZF domains per species (Figure 1B, col. 4) that are used for comparative analysis in further steps of our framework.
  
  \subsection*{Substantial Loss and Recruitment of C2H2-ZF Domains Relative to D. \\ melanogaster}
  We first assessed the loss and gain of C2H2-ZF domains across our orthogroups, as the number and arrangement of C2H2-ZF domains likely affects the binding specificity of each poly-ZF gene. \emph{D. melanogaster} domains are considered ``lost'' in each non-reference species without a corresponding aligned domain; \emph{D. melanogaster} domains with no aligned domains in any of the other fly species are ignored because they most likely are species-specific \emph{D. melanogaster} gains. Conversely, domains from non-\emph{melanogaster} sequences that did not align back to a \emph{D. melanogaster} domain are considered ``gains'' with respect to the reference.
  
  Between 1.4\% and 8.9\% of \emph{D. melanogaster} domains were lost in the other fly species (Figure 2A), and between 0.5\% and 1.8\% of domains from non-\emph{melanogaster} species were gained with respect to the reference (Figure 2B). A notable 19.6\% of all non-reference poly-ZF genes in 1-to-1 orthologous relationships with a \emph{D. melanogaster} gene have lost or gained a C2H2-ZF domain with respect to the reference. 75.6\% of gains or losses occur outside of or at an end of an array of canonically linked domains. The proportion of domains lost and gained in the non-\emph{melanogaster} species with respect to the reference increases as the phylogenetic distance from \emph{D. melanogaster} increases. When considering only canonically linked C2H2-ZF domains, we still see the same overall phylogenetic trends, albeit at a somewhat lower level.
  
  We note that \emph{D. melanogaster} benefits from more complete sequencing coverage in comparison to the the other fly genomes \cite{drosophila2007}, and relatively poor coverage and subsequent inaccurate sequence assembly would result in a greater number of unidentified or misidentified domains in those genomes. \emph{D. sechellia}, \emph{D. simulans}, and \emph{D. persimilis}, which exhibit the greatest relative C2H2-ZF domain loss (Figure 2A), also have the lowest relative coverage: 4.9x, 2.1x, and 4.1x, respectively, compared to between 8.4x and 11.0x for the other species. For this reason, the C2H2-ZF domain gains relative to \emph{D. melanogaster} are especially noteworthy, while some of the apparent domain losses, especially from \emph{D. sechellia}, \emph{D. simulans}, and \emph{D. persimilis}, may be due to incomplete assemblies.

  \subsection*{Pervasive Variation in Specificity-Determining Residues in Aligned C2H2-ZF Domains}
  Binding specificity may also be altered as a result of deviations in the DNA-contacting, specificity conferring residues in positions -1, 2, 3, or 6 of the C2H2-ZF domain \cite{wolfe2000} (Figure 3A).  With the exception of structurally constrained position 4 within the domain's alpha-helix, these functional sites are more conserved than their neighboring residues, yet still show comparably high divergence (Figure 3B).  We consider an aligned domain in any non-reference fly species to be ``diverged'' if at least one of its residues from positions -1, 2, 3, or 6 has diverged from the \emph{D. melanogaster} reference.  Of the $>$98\% of domains from poly-ZF genes that aligned between the non-reference sequences and their orthologs in \emph{D. melanogaster}, we observe from 6.9\% domains diverged in \emph{D. sechellia} (last common ancester [LCA] with \emph{D. melanogaster} $\sim$2 Mya) to a substantial 36.4\% domains diverged in \emph{D. mojavensis} (LCA with \emph{D. melanogaster} $\sim$45 Mya) (Figure 3C).  These divergent domains are not confined to a small subset of genes: across the 11 non-reference fly species, 11.1\% to 50.0\% of poly-ZF genes with 1-to-1 orthologs in \emph{D. melanogaster} contain at least one divergent C2H2-ZF domain. Moreover, as with the proportion of domains lost and gained with respect to the reference, the proportion of domains diverged steadily increases as phylogenetic distance from \emph{D. melanogaster} increases, and the same trend with slightly lower overall divergence is observed in the subset of canonically linked domains.  Of the 41.7\% of domains situated in the middle of canonically linked arrays, 18.2\% contain divergent binding residues; of the remaining domains outside of or flanking canonically linked arrays, 27.2\% contain divergent binding residues. Arrays of canonically linked domains appear to be under stricter constraints than singleton domains are (Figure S2A-C). Altogether, changes in these DNA-contacting residues are substantially more frequent than the complete gain or loss of C2H2-ZF domains.

  \subsection*{Functional and Evolutionary Importance of Divergences}
    \subsubsection*{Diverging DNA-binding residues show conservation within phylogenetic clades}
    We reasoned that changes in these single-copy TFs relative to \emph{D. melanogaster} that are functionally important are likely to be conserved across phylogenetic clades. To test this, we extracted pairs of sequences from the most closely related species --- \emph{D. sechellia} and \emph{D. simulans} (LCA $<$2 Mya), \emph{D. pseudoobscura} and \emph{D. persimilis} (LCA $\sim$2 Mya), \emph{D. yakuba} and \emph{D. erecta} (LCA $\sim$5 Mya), and \emph{D. virilis} and \emph{D. mojavensis} (LCA $\sim$25 Mya) --- and asked how often a particular mutation with respect to the reference in one species was supported by an identical mutation in its partner species. In all cases, divergent binding residues in poly-ZF genes from each non-\emph{melanogaster} fly species exhibit clade support more often than other background divergent residues in these genes (Figure 4A and Figure 4B, col. 1); these changes are significant ($p$ {\tt{<}} 0.001, binomial test) in 4 species, with small sample sizes a limiting factor in the other species (Table S1). Residues within and between adjacent C2H2-ZFs are also under structural constraints, resulting in their high conservation according to the clade support measure, particularly in the species closest to \emph{D. melanogaster} with fewer overall divergent residues. Altogether, this analysis suggests that the substantial binding residue variation we see across species is functional rather than random.

    \subsubsection*{Relatively low evolutionary rate of diverging DNA-binding residues}
    To further support the claim that observed variations in C2H2-ZF binding residues are functionally important, we estimate site-based evolutionary rates using Rate4Site \cite{rate4site} for all divergent residues per sequence per orthogroup (Figure 4B, col. 2). For each sequence, we ranked its divergent residues from lowest to highest evolutionary rates, and normalized these ranks to values between 0 and 1. In each non-reference species across all orthogroups, divergent binding residues evolve slower than background residues outside of C2H2-ZF domains ($p$ {\tt{<}} 0.001 in the 10 species furthest from \emph{D. melanogaster}, Wilcoxon test; Table S1). Because we consider only divergent residues in each sequence, this signal is strongest in species with a large number of total divergent residues per sequence, as normalized ranks are more continuous in these cases and therefore differences between the four classes of residues (i.e. specificity-conferring binding residues, background, non-helical C2H2 residues, and linker regions) are apparent with higher resolution. In the flies furthest away from \emph{D. melanogaster}, where the most variation from \emph{D. melanogaster} is observed, the divergent binding residues exhibit the slowest evolutionary rate relative to all other classes of divergent residues, including the structurally constrained non-binding regions within domains and the linker regions between domains in poly-ZF genes. In the four species closest to \emph{D. melanogaster}, non-binding residues in C2H2-ZF domains appear to evolve as slowly as binding residues themselves, as the low number of total divergent residues per sequence (Table S1) restricts the resolution of differences between the residue classes.
    
    \subsubsection*{Population analysis suggests positive selection in evolutionary history}
    We have shown that divergent binding residues are under functional constraints, yet the pervasiveness of such changes in these 1-to-1 orthologs suggests these deviations may confer an evolutionary advantage and were selected for when they arose. The classic approach for detecting positive selection is to calculate dN/dS, the ratio of observed non-synonymous mutations over all possible non-synonymously mutable sites to observed synonymous mutations over all possible synonymously mutable sites \cite{kimura1977}. However, we found that across the \emph{Drosophila} species, dS naturally saturates and thus impedes a positive selection signal of dN/dS $>$ 1, suggesting that this measure is inappropriate for use across the  evolutionary distances considered here \cite{yang1997}.
    
    In order to detect positive selection in the evolutionarily dispersed fly species, therefore, we utilize an alternate, population based approach.  If a nonsynonymous mutation was neutral and became fixed via random genetic drift, it is more likely to persist as a polymorphism within a population, whereas if such a mutation was advantageous and became fixed rapidly through positive selection, finding nonsynonymous mutations in the same location in population data would be highly unlikely \cite{mcdonald1991}.  Consequently, for each divergent residue in each non-reference species, we asked whether that same site was or was not polymorphic in a population of 139 \emph{D. melanogaster} organisms \cite{pool2012} (Figure 4B, col. 3).  In every species, a greater proportion of divergent binding residues are disjoint from polymorphic sites than divergent background residues are ($p$ {\tt{<}} 1e-15 in all species, binomial test; Table S1), and in 7 of the 11 species these proportions are greater than those for all other types of diverging residues.  In four species, linker regions between domains, which may impact overall specificity by affecting flexibility of canonical binding arrays and the positioning of C2H2-ZF domains within them, had residue changes present as polymorphisms as often as the binding residues themselves. In two species, diverging non-helical residues, which may alter the structure of the DNA-binding domains, also overlapped with polymorphic sites in \emph{D. melanogaster} as rarely as binding residues did. Altogether, this evidence suggests that a substantial proportion of binding residue divergences in each species, therefore, were likely advantageous rather than neutral compared with other variation within poly-ZF genes.
    
  \subsection*{Divergent Residues Lead to Distinct Computationally-Predicted Specificities}
    Despite the functionally relevant sequence divergence we see in C2H2-ZF domains, it is possible for distinct assignments of binding residues to still specify the same overall recognition motif. As a result, we next aimed to ascertain how the variation we observe in poly-ZF orthologs may change DNA-binding specificity.  We predicted the specificity of each C2H2-ZF domain with a predictor \cite{persikov2014} that utilizes a linear support vector machine based on an expanded structural model (Figure 3A); this method is referred to as SVM. Since no method can predict binding specificity perfectly, we compared the SVM predictions to those produced by an independent predictor referred to as ML that utilizes a probabilistic recognition code generated via maximum likelihood \cite{benos2002}, and a random forest based predictor referred to as RF \cite{gupta2014}. We calculated the average Pearson correlation coefficients (PCCs) across positions b1 through b4 between SVM predicted position weight matrices (PWMs) and ML and RF PWMs, and consider only the subset of SVM predictions with average PCCs greater than 0.25 to either of the corresponding ML or RF predictions (Figure S3).  Of the 17319 aligned binding domains from all 12 fly species, 87.2\% passed this confidence threshold. Results using alternate confidence thresholds of PCC $>$ 0.0, PCC $>$ 0.5 and PCC $>$ 0.75 are found in Table S2.
    
    We compare the SVM-predicted PWM for each divergent domain in a non-\emph{melanogaster} species to the predicted PWM for its aligned domain in its \emph{D. melanogaster} ortholog by calculating the average PCC across positions b1 through b4. In six non-reference fly species, 100\% of all divergent domains exhibit a PCC $<$ 1 from their reference domains in at least one predicted position (Figure S4A). Of the remaining five species, $<$1\% of divergent domains do not show a significant change in predicted specificity in any position compared to their aligned \emph{D. melanogaster} reference domains. Many domains from non-\emph{melanogaster} species exhibit a diverged specificity from the reference in more than one predicted position (Figure S4B). In each species, the same proportion of divergent canonically linked domains exhibit a change in predicted specificity from the reference (data not shown). This shows that the divergent binding residues within C2H2-ZF domains likely result in changed DNA-binding specificity.
    
    \subsubsection*{Binding Specificities Change Gradually over Evolutionary Distance}
    To establish how specificity may change in relation to phylogenetic distance, we compare the predicted PWM for each \emph{D. melanogaster} domain to those PWMs predicted for every corresponding aligned domain from the other flies (example domain multiple alignment in Figure 5A-B).  In a majority of such domain alignments between \emph{D. melanogaster} and the other fly species, we note that as the phylogenetic distance between \emph{D. melanogaster} and the other fly species increases, the PCCs comparing binding specificities between \emph{D. melanogaster} domains and their non-reference orthologs decreases (Figure 5C-E). For each domain alignment, the phylogenetic (species) distance of each domain from \emph{D. melanogaster} is negatively correlated to the amount of its predicted specificity change from \emph{D. melanogaster}, with a Spearman correlation $<$ 0 for 88.9\% of domain alignments and $< -$0.5 for 65.3\% of domain alignment (see Table S2). When we group divergent non-\emph{melanogaster} domains by species rather than by orthology to a particular \emph{D. melanogaster} domain, we still observe this same negative trend, where an increase in phylogenetic distance between \emph{D. melanogaster} and non-reference species correlates with a decrease in the PCCs between predicted \emph{D. melanogaster} specificities and non-reference predicted specificities (Figure 5F). Overall, our analysis of predict specificities suggests that DNA-binding specificities can diverge gradually over evolutionary time in non-duplicated, 1-to-1 C2H2-ZF orthologs.
    
    \subsection*{Diverged Poly-ZFs are Functionally Varied; Conserved Poly-ZFs are Developmentally Enriched}
    Do divergent poly-ZF genes exhibit distinct biological functions from the set of conserved poly-ZF genes? To answer this question, we divided the genes from our analysis into two sets. The first set contained 73 poly-ZF genes from \emph{D. melanogaster} with  completely conserved DNA-contacting residues across all its orthologs; 27 (36.5\%) had orthologs in all other fly species, and 65 (87.8\%) contained canonically linked domains. The second set contained 161 \emph{D. melanogaster} poly-ZF genes with a diverged C2H2-ZF domain in 2{\tt{+}} orthologs; 76 (47.2\%) had orthologs in all 11 other fly species, 145 (90.1\%) contained canonically linked domains, and 135 (83.9\%) contained a divergent canonically linked domain.
  
    \subsubsection*{Divergent Poly-ZFs have Limited Functional Annotations}
    We ran GO Term Finder \cite{boyle2004} on these two gene sets to find enrichment of Gene Ontology terms (Table S3). Both the conserved and divergent sets are separately enriched for {\sf DNA-templated regulation of transcription}, {\sf positive or negative regulation of gene expression}, and {\sf regulation of RNA metabolic process} ($p$ {\tt{<}} 0.001, Bonferroni-corrected hypergeometric test), even when excluding annotations inferred from sequence models. Unsurprisingly, those poly-ZF genes with conserved binding specificities are also enriched for such developmental functions as {\sf segmentation}, {\sf morphogenesis}, and {\sf organ development}. The poly-ZF genes with divergent C2H2-ZF domains, on the other hand, exhibit no additional functional enrichments, even when considering only genes with orthologs in every species or with divergent canonically linked domains. Although no functions were significantly enriched for across the entire set of divergent poly-ZF genes, certain genes within this set were annotated with functions such as {\sf organ development (muscle, respiratory system, axon, wing disc)}, {\sf dorsal/ventral pattern formation}, and {\sf neurogenesis}. These results highlight a clear study bias toward conserved, developmentally involved TFs.

    \subsubsection*{Co-Domain Presence Suggests Trancriptional Regulation Activity}
    To get a better sense of these genes' functions, we also downloaded domain annotations from InterPro \cite{interpro} to discover co-domain overlaps between the sets of conserved and diverged poly-ZF genes. Both of these gene sets contain the regulation related effector domain {BTB/POZ}, which mediates homomeric dimerization, and additional DNA-binding {AT-hook} and {homeobox} domains. The domains found uniquely in the conserved set are {DZF}, a nucleotidyltransferase; {SET}, a histone methyltransferase found predominantly in enhancer TFs; {Ovo}, which plays a role in germline sex determination; {SANT/Myb}, another DNA-binding domain; and {ELM2}, a domain of unknown function. Divergent C2H2-ZF genes uniquely contain several domains implicating their regulatory activity --- {PHD}, responsible for chromatin-mediated transcriptional regulation; {PWWP}, {ING}, {WD40}, and {bromodomain}, all important for chromatin remodeling, genome stability maintenance, protein-histone association, and cell cycle progression regulation; {EPL1}, involved in transcriptional activation; and {BESS}, {TRAF}, and {SWR1}, which direct a variety of protein-protein interactions. 

    \subsubsection*{Cases of Specific TFs with Divergent C2H2-ZF Domains}
    Despite the limited availability of experimental data about C2H2-ZF genes with varying DNA-binding domains, several known TFs are found in the set of divergent genes. For instance, Matotopetli (\emph{topi}), a testis-specific regulator of meiosis and terminal differentiation \cite{perezgasga2004}, has 11 C2H2-ZF domains in \emph{D. melanogaster}, of which five were mutated in the six species furthest from \emph{D. melanogaster}, four were mutated in \emph{D. ananassae}, two were mutated in \emph{D. yakuba}, and one was mutated in \emph{D. sechellia} and \emph{D. erecta}. Tiptop (\emph{tio}), a repressor of the teashirt TF and regulator of clypeolabrum patterning \cite{laugier2005}, has five C2H2-ZF domains, the first, third, and fifth of which have diverged from the \emph{D. melanogaster} ortholog in seven other species. Hermaphrodite (\emph{her}), a regulator required for sexual differentiation \cite{li1998}, has four C2H2-ZFs, the first of which has a mutation in position 2 in \emph{D. yakuba} and \emph{D. erecta}, the fourth of which has a mutation in position 2 in \emph{D. pseudoobscura} and \emph{D. persimilis}, and the second and third of which both have mutations in position 2 in \emph{D. willistoni}. Evidence for DNA binding for other poly-ZF genes with divergent domains comes from the experimentally-determined DNA-binding specificities for 58 distinct \emph{D. melanogaster} poly-ZF genes included in our analysis \cite{flyfactorsurvey,enuameh2013}; this set includes 22 poly-ZF genes that each contain at least one domain that diverges across the flies.
    
    \subsubsection*{Divergent Poly-ZFs are less essential and more widespread}
    Additional phenotypic information derived from gene knockout experiments are available via FlyBase and modENCODE \cite{modencode} for 97.3\% of conserved poly-ZF genes and 96.2\% of divergent poly-ZF genes. Conserved poly-ZF genes are more often essential than divergent poly-ZF genes are: gene knockouts were lethal for 23.3\% of conserved and only 15.8\% of diverged genes. An additional 43.8\% and 21.5\% of conserved and divergent genes respectively had semi-lethal, recessive lethal, or larval lethal knockouts. In concurrence with the GO term enrichment, we found that 37.0\% of conserved poly-ZF gene knockouts affected phenotypes in the embryonic or larval stages, whereas only 12.0\% of diverged poly-ZF knockouts had a phenotypic effect during development.  Only 8 of the conserved and 3 of the diverged \emph{D. melanogaster} poly-ZFs in our analysis had associated ChIP binding site information available from modENCODE. We uncovered no significant differences between the two sets with regard to count or location of their \emph{in vivo} binding sites.
    
    To further determine where and when poly-ZF genes affect phenotype, we looked at expression locale and levels derived from FlyAtlas \cite{flyatlas}, available for 95.8\% of conserved poly-ZFs and 97.5\% of diverged poly-ZFs. We considered adult, larval, and germline tissues separately (Figure S5A). Interestingly, we found that larger proportions of divergent poly-ZFs were found in each tissue than the proportions of conserved poly-ZFs.  Although divergent poly-ZFs tended to be present in a larger number of distinct tissues than conserved poly-ZFs were (Figure S5B), their expression was consistently lower than the expression of conserved poly-ZFs in corresponding tissues (Figure S5C).  

  \section*{Discussion}

  Previously, binding site turnover has been shown via ChIP experiments to be an essential component in regulatory network variation across closely-related organisms \cite{tuch2008,borneman2007,bradley2010,odom2007,schmidt2010} and even across individuals of the same species \cite{kasowski2010,zheng2010}. Here we present an analysis suggesting that divergence of orthologous TFs also plays a role in regulatory variation.

  Over half of the single-copy, poly-ZF 1-to-1 gene orthogroups in \emph{Drosophila} exhibit variation with respect to the number and arrangement of DNA-binding C2H2-ZF domains and the composition of specificity-conferring residues within these domains. Structural models of C2H2-ZF--DNA complexes suggest that such mutations influence the binding specificity of the genes in which they are found.  These mutations' conservation across phylogenetic clades, low rate of evolution, and rapid fixation as determined by their lack of overlap with population SNPs further demonstrate their functional importance. Additionally, predicted specificities of C2H2-ZF domains increasingly diverge as evolutionary distance from the reference \emph{D. melanogaster} increases, offering evidence that specificity-altering \emph{trans} changes are feasible and occur in evolutionarily viable steps even in non-duplicated orthologs.

  Though C2H2-ZF binding of RNA \cite{pelham1980} or protein \cite{brayer2008} rather than or in addition to DNA has been observed, several lines of evidence suggest that a large fraction of the domains in our study are binding DNA. We focus on only those genes with multiple C2H2-ZF domains, a requirement for specific DNA recognition. Even when we limit our analysis to canonically linked domains, which have the strongest structural evidence for DNA-binding, we observe the same overall divergence trends. Some DNA-binding C2H2-ZFs may regulate processes other than transcription; however, GO term enrichment analysis and co-domain presence suggests that many of these poly-ZFs are regulating transcription and gene expression and are likely interacting with other protein co-factors. Altogether, this suggests that a substantial set of the divergent poly-ZF genes included in our analysis are DNA-binding TFs.  However, it is also possible that the likely specificity-altering mutations we see in these DNA-binding TFs may leave overall gene expression unaffected. There are cases of divergent \emph{cis}-regulatory sequences that do not confer a change in gene expression \cite{dermitzakis2002,costas2003,moses2006,doniger2007,kim2009,venkataram2010}, review by \cite{weirauch2010}, as sometimes these binding site changes are accompanied by complementary TF changes \cite{tan2008}.  Compensatory change may occur for some of the diverging poly-ZF TFs we observe. Nevertheless, the substantial \emph{trans} variations probably result in, at minimum, modulated expression changes, as multiple \emph{cis} mutations co-occurring with and counteracting each \emph{trans} specificity change would be extremely unlikely.  
  
  Although prior research has recognized the possibility of TF variation occurring in multi-gene families, it has long been thought that single-copy TFs are under stringent conservation, as loss or change of function mutations in these genes could not be masked by the functional gene products of paralogs and would thus have catastrophic effects. Gene duplication was thought to be the primary way by which sub- or neofunctionalization of TFs may occur, as aggregate gene copies could still maintain the original gene's function.  However, our large-scale results on 1-to-1 C2H2-ZF orthogroups in flies are consistent with a recent experimental case study of specificity divergence of a single-copy TF in plants \cite{sayou2014}.  Here, binding specificities of 1-to-1 orthologs of the plant TF LEAFY (\emph{lfy}) were analyzed across nine algal, moss, and plant species, and three distinct binding preferences were found. One of the \emph{lfy} orthologs was dubbed a ``promiscuous intermediate'' as it recognizes all three binding motifs with various preferences. This intermediate highlights a means by which TF binding specificity can evolve in single-copy genes which have not undergone a prerequisite gene duplication event. The gradual TF variation we observe may also give rise to such analogous TF intermediates.
  
  In conclusion, we propose that variation in 1-to-1 orthologous TFs can also shape regulatory network evolution. Changes in TFs need not be catastrophic. Rather, single amino acid mutations in DNA- contacting positions may result in overall TF binding of similar targets with varying affinities. Such variations provide the opportunity for gradual evolution of binding specificity. We propose that these changes in single-copy TFs may be substantial contributors to overall regulatory evolution in \emph{Drosophila} and in other metazoans in general.

\section*{Methods}

\subsection*{M1. Sequence Collection}
    Translated protein sequences for the 12 sequenced fly species --- \emph{D. melanogaster} (build r5.53), \emph{D. sechellia} (r1.3), \emph{D. simulans} (r1.4), \emph{D. yakuba} (r1.3), \emph{D. erecta} (r1.3), \emph{D. ananassae} (r1.3), \emph{D. pseudoobscura} (r3.1), \emph{D. persimilis} (r1.3), \emph{D. willistoni} (r1.3), \emph{D. mojavensis} (r1.3), \emph{D. virillis} (r1.2), and \emph{D. grimshawi} (r1.3) --- were downloaded from FlyBase \cite{flybase}. Additional \emph{D. simulans} sequences were downloaded from the Andolfatto Lab site \cite{hu2013}. To identify C2H2-ZF genes, HMMER's {\tt{hmmsearch}} (versions 2.3.2 \cite{hmmer} and 3.0 \cite{hmmer3}) was run on each translated protein file using 12 Pfam HMMs \cite{pfam}, which were selected based upon their similarity to and presence in the same clan as the consensus Cys$_2$-His$_2$ zinc finger profile (Figure S1B), zf-C2H2 (PF00096) --- zf-C2H2 (PF00096), zf-C2H2\_2 (PF12756), zf-C2H2\_6 (PF13912), zf-C2H2\_jaz (PF12171), zf-C2HC\_2 (PF13913), zf-H2C2\_5 (PF13909), zf-met (PF12874), zf-met2 (PF12907), zf-BED (PF02892), zf-U1 (PF06220), GAGA (PF09237), DUF3449 (PF11931). Any protein sequence containing at least one HMMER hit with a bit score above the specified gathering domain threshold for that HMM was considered.
    
    C2H2-ZF domains themselves were identified from these proteins as any HMMER hit matching the regular expression {\tt{CX$_{2,}$CX$_{8,}\Psi$X$_2$HX$_{3,}$[H$|$C]}}, where $\Psi$ is a large, hydrophobic amino acid. Hits that did not match this expression and thus no longer have the structure necessary to bind DNA are considered degenerate, and are not identified as domains. HMMER hits below the corresponding bitscore thresholds but which matched this regular expression were retained in these proteins because C2H2-ZFs are known to occur in tandem, and therefore we are more confident about all C2H2-ZF domains which co-occur with at least one high scoring domain.
    
    Where possible, the longest protein splice form per gene containing a superset of all C2H2-ZF domains was selected to represent each gene. If no single protein isoform contained a superset of domains, a minimal set of proteins which together include all unique C2H2-ZF domains was selected to represent the gene.

  \subsection*{M2. Orthogroup Collection \& Augmentation}
    A list of pairwise orthologs to \emph{D. melanogaster} was downloaded from FlyBase and from the Andolfatto Lab build of \emph{D. simulans} \cite{hu2013}, and orthogroups were constructed from overlaps of these orthologs. Those orthogroups containing at least one \emph{D. melanogaster} poly-ZF gene were selected. Of 13172 total orthogroups, 272 had at least one \emph{D. melanogaster} poly-ZF gene.
	
    \emph{D. melanogaster} poly-ZF orthogroups with sequences missing from one or more species were augmented according to the 15 insect whole genome alignment (WGA) from the UCSC Genome Browser \cite{ucscgb}. A missing species is defined as any species not present in the orthogroup but present in the phylogenetic subtree rooted at the most recent common ancestor of those species that are present in the orthogroup. For each of the 36.1\% of orthogroups containing at least one missing species, known protein sequences were aligned to the UCSC 15-insect WGA using BLAT \cite{blat}. Where possible, sequence(s) from the missing species were extracted from the section of the alignment with the best hits and aligned back to their corresponding translated protein files using BLAT again. Gene IDs of proteins with BLAT hits with an e-value cutoff of 0.001 were extracted and, when they were not present in pseudogene lists, were added to the corresponding orthogroups. Through this process, 51\% of the orthogroups with missing species were augmented with at least one new gene. 
	
  \subsection*{M3. Orthogroup Reconciliation}
    All 1-to-many (i.e. one gene from \emph{D. melanogaster} but more than one gene from at least one other species) orthogroups were truncated such that only those species with a single gene in the original orthogroup were included in the new orthogroup. In this manner, our analysis was restricted to variation in 1-to-1 orthologs.
	
    A gene tree was constructed for each many-to-many orthogroup using T-Coffee \cite{tcoffee}, and each gene tree was reconciled with the phylogenetic species tree using Notung \cite{notung} to identify duplication and loss events. Genes that were 1-to-1 with a non-lost \emph{D. melanogaster} gene in subtrees of the reconciled trees were extracted as new 1-to-1 orthogroups; genes that did not fit this description were discarded.

\section*{Acknowledgments}

  We thank other members of the Singh lab for their insight and comments. Thanks in particular to Jesse Farnham and Dario Ghersi for their technical assistance with phylogenetic analysis using Notung and gene construction from nucleotide sequencing files, and to Yuri Pritykin for his compiled fly expression data parsed from FlyAtlas. We also thank Peter Andolfatto for a very helpful discussion about this work. This work was funded in part by NIH grant R01-GM076275 (to MS) and NSF GRFP grant DGE-1148900 (to SN).

\section*{Author Contributions}
  MS, SN and AP designed the study. SN and AP performed the analysis. SN drafted the initial manuscript. All authors read and approved the final manuscript.

\bibliographystyle{plos2009}
 \bibliography{NadimpalliZF}
 
\newpage 
 
\section*{Figures}

\begin{adjustwidth}{-1in}{-1in}
  \begin{center}
    \includegraphics{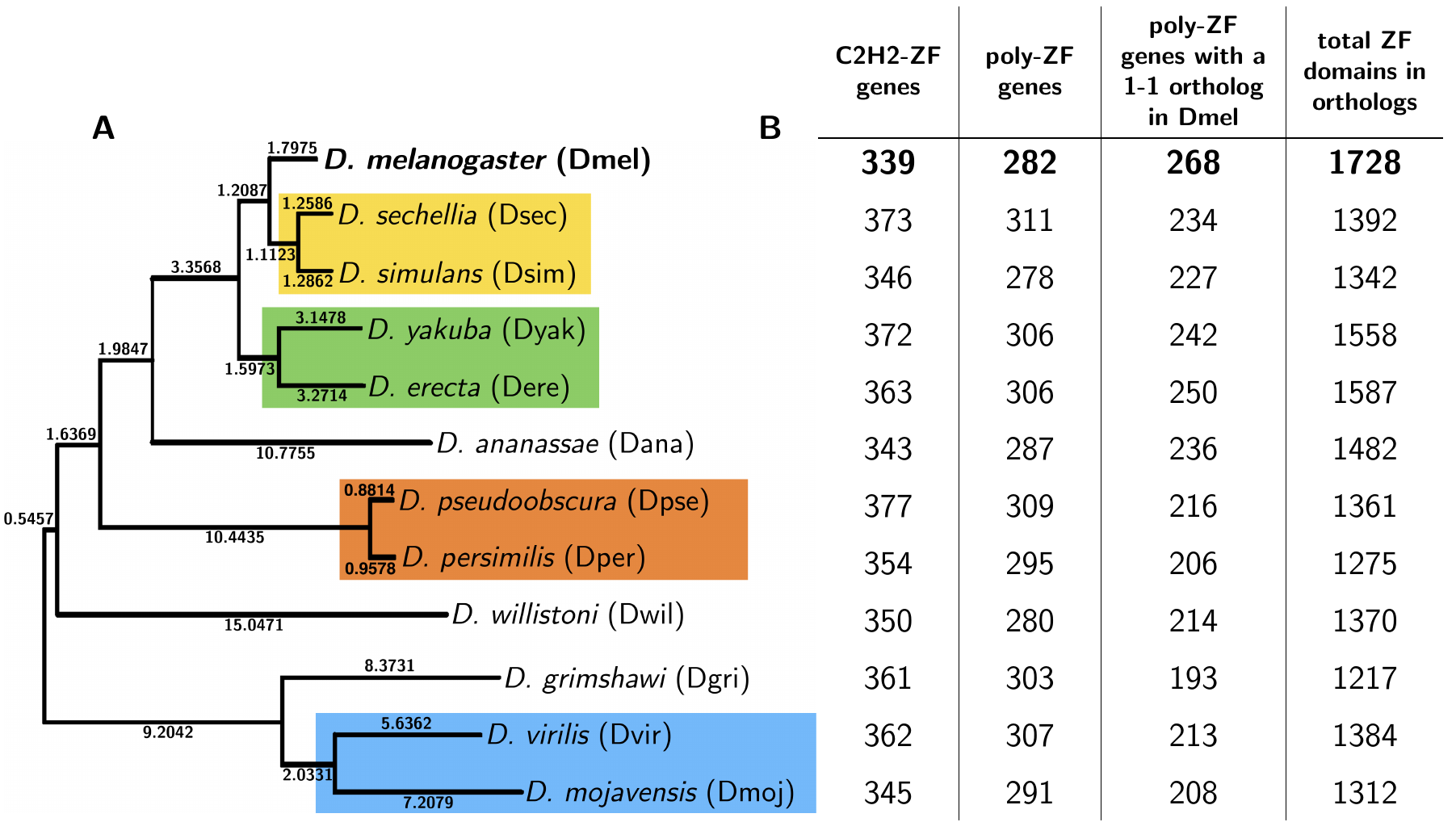}
  \end{center}
\end{adjustwidth}

\vspace{1cm}
\noindent
{\bf Figure 1. Phylogenetic tree relating 12 \emph{Drosophila} species.} 
\\
(A) Phylogenetic tree of 12 \emph{Drosophila} species. The four-letter abbreviations of the species are given with \emph{D. melanogaster}, the reference sequence, in bold. Branch lengths are as reported in the UCSC Genome Browser. Pairs of the most closely related species are highlighted in colored boxes. (B) Columns correspond to counts of, for each species: C2H2-ZF genes, C2H2-ZF genes with 2{\tt{+}} C2H2-ZF domains (poly-ZF genes), poly-ZF genes with a 1-to-1 ortholog in \emph{D. melanogaster}, and the number of domains found in the set of poly-ZF genes with a 1-to-1 ortholog in \emph{D. melanogaster}. In the \emph{D. melanogaster} row, the last two columns correspond to the count of poly-ZF genes with a 1-to-1 ortholog in any of the other 11 fly species and the number of domains found in this set of poly-ZF genes.

\newpage

\begin{adjustwidth}{-1in}{-1in}
  \begin{center}
    \includegraphics{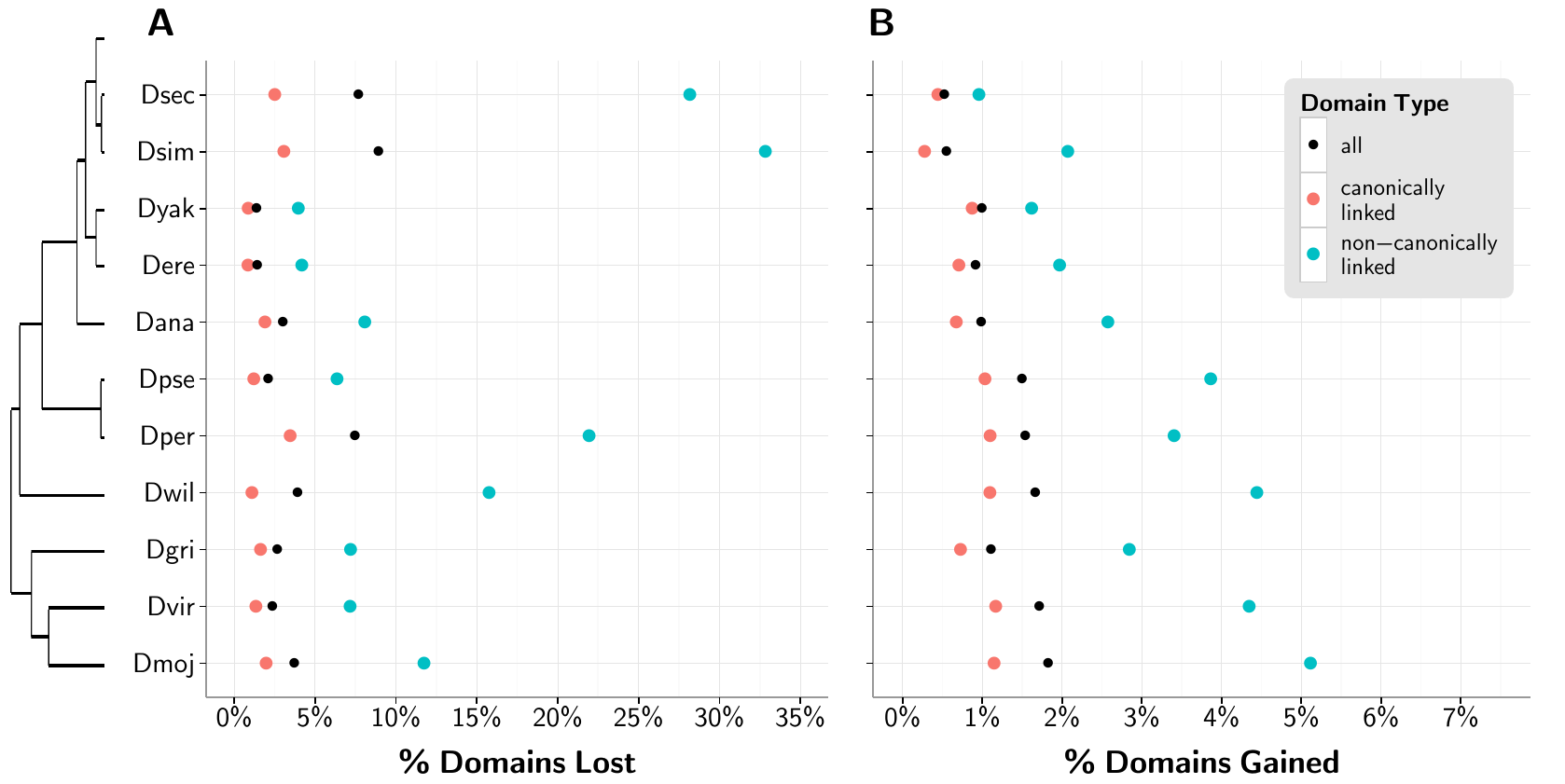}
  \end{center}
\end{adjustwidth}

\vspace{1cm}
\noindent
{\bf Figure 2. Loss and recruitment of C2H2-ZF domains with respect to \emph{D. melanogaster} reference.}
\\
      (A) Percent loss of \emph{D. melanogaster} domains in each non-reference species for all domains (black) and separately for non-canonically linked (blue) and canonically linked (red) domains, with a phylogenetic tree relating the fly species to the left. (B) Percent domains gained by each non-\emph{melanogaster} species. 

\newpage

\begin{adjustwidth}{-1in}{-1in}
  \begin{center}
    \includegraphics{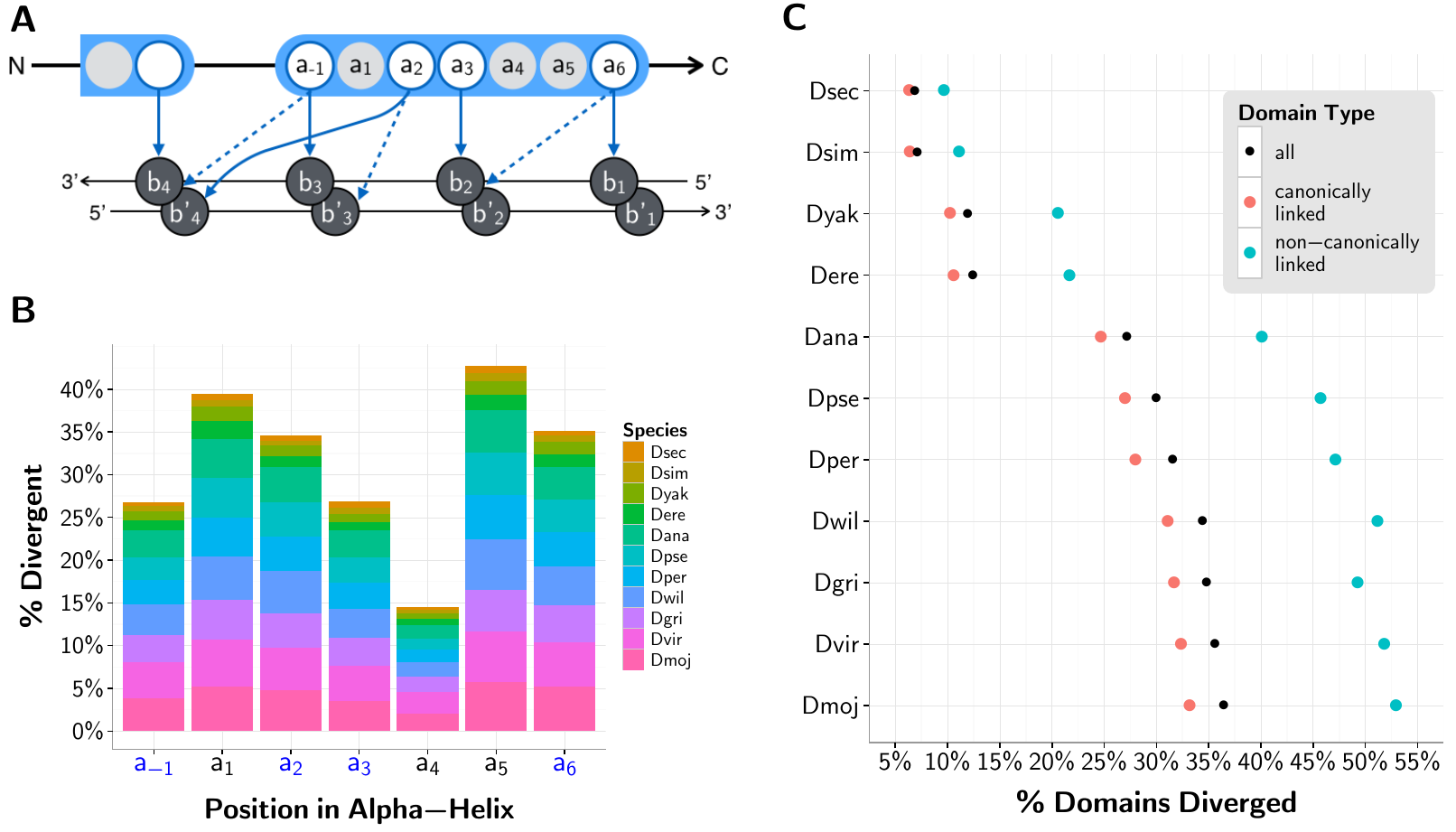}
  \end{center}
\end{adjustwidth}

\vspace{1cm}
\noindent
{\bf Figure 3. C2H2-ZF domain divergence with respect to \emph{D. melanogaster} reference.}
\\
    (A) Schematic of a C2H2-ZF protein--DNA interface under the 7-contact model \cite{persikov2011}. Amino acids within the depicted finger are numbered according to their relative position from the start of the alpha-helical domain, with $a_{–1}$ denoting the residue immediately preceding the helix. Bases $b_1$, $b_2$, $b_3$ and $b_4$ are numbered sequentially from 5' to 3' of the primary DNA strand; the complementary bases are denoted by b1', b2', b3' and b4'. Contacts between amino acids and bases are shown in arrows, with four specificity-determining amino acids $a_{-1}$, $a_2$, $a_3$ and $a_6$ making these contacts. (B) Histogram showing the percent divergence per species by position within the C2H2-ZF domain's alpha-helix (-1 to 6) for all canonically linked domains. The columns with blue labels in the $x$-axis correspond to positions that interact with DNA in the 7-contact model.  (C) Percent of all (black), canonically linked (blue) and non-canonically linked (red) aligned domains in each non-reference fly species with a divergent residue (as compared to the \emph{D. melanogaster} reference) in positions -1, 2, 3, and/or 6. 

\newpage

\begin{adjustwidth}{-1in}{-1in}
  \begin{center}
    \includegraphics{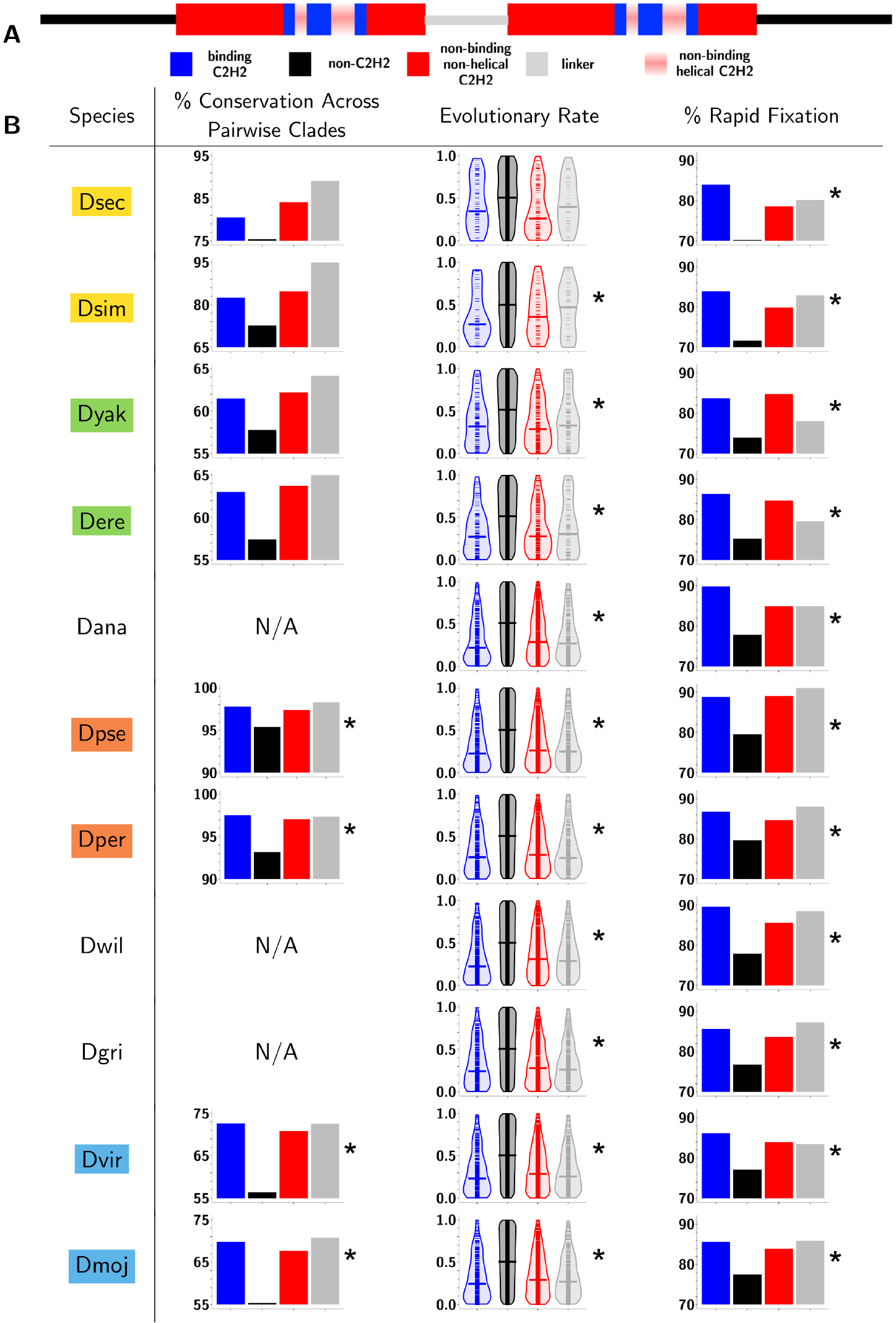}
  \end{center}
\end{adjustwidth}

\vspace{1cm}
\noindent
{\bf Figure 4. Functional importance of C2H2-ZF gene residues.}
\\
      (A) Legend depicting a sequence with non-C2H2-ZF domain residues (black), residues in non-binding regions of the C2H2-ZF domain outside of the alpha-helix (red), the four DNA-binding residues in the alpha-helix (blue), and linker regions between adjacent canonically linked C2H2-ZF domains (gray). Positions 1, 4, and 5 in the alpha-helix (pink) are not included in the analysis because while they are typically not DNA binding, they are found in the recognition helix. (B) Values for each of these four residue classes for the following: percent of divergent residues with a matching mutation in the most closely paired species (left); ranks of divergent residues based on evolutionary rate as predicted by Rate4Site \cite{rate4site} that have been 0-to-1 normalized (middle); and percent of residues found to diverge between \emph{D. melanogaster} and each other fly species that did \emph{not} correspond to a polymorphic site in \emph{D. melanogaster} population data (right). The plots in the middle column are violin plots, where the dynamic widths of the violins correspond to the relative density of points in the distribution, and the medians are given by horizontal lines. The areas of the violins are equal per plot. Plots depicting significant ($p$ {\tt{<}} 0.001) differences between the binding residues (blue) and background residues (black) are marked with an asterisk to the right of the plot. The $p$-values, calculated using a binomial test in the left and right columns and wilcoxon test in the middle column, range from 10$^{-3}$ to 10$^{-21}$ (left), 10$^{-3}$ to 10$^{-55}$ (middle) and 10$^{-15}$ to 10$^{-124}$ (right). Exact $p$-values for all plots can be found in Table S1.

\newpage

\begin{adjustwidth}{-1in}{-1in}
  \begin{center}
    \includegraphics[width=6.5in]{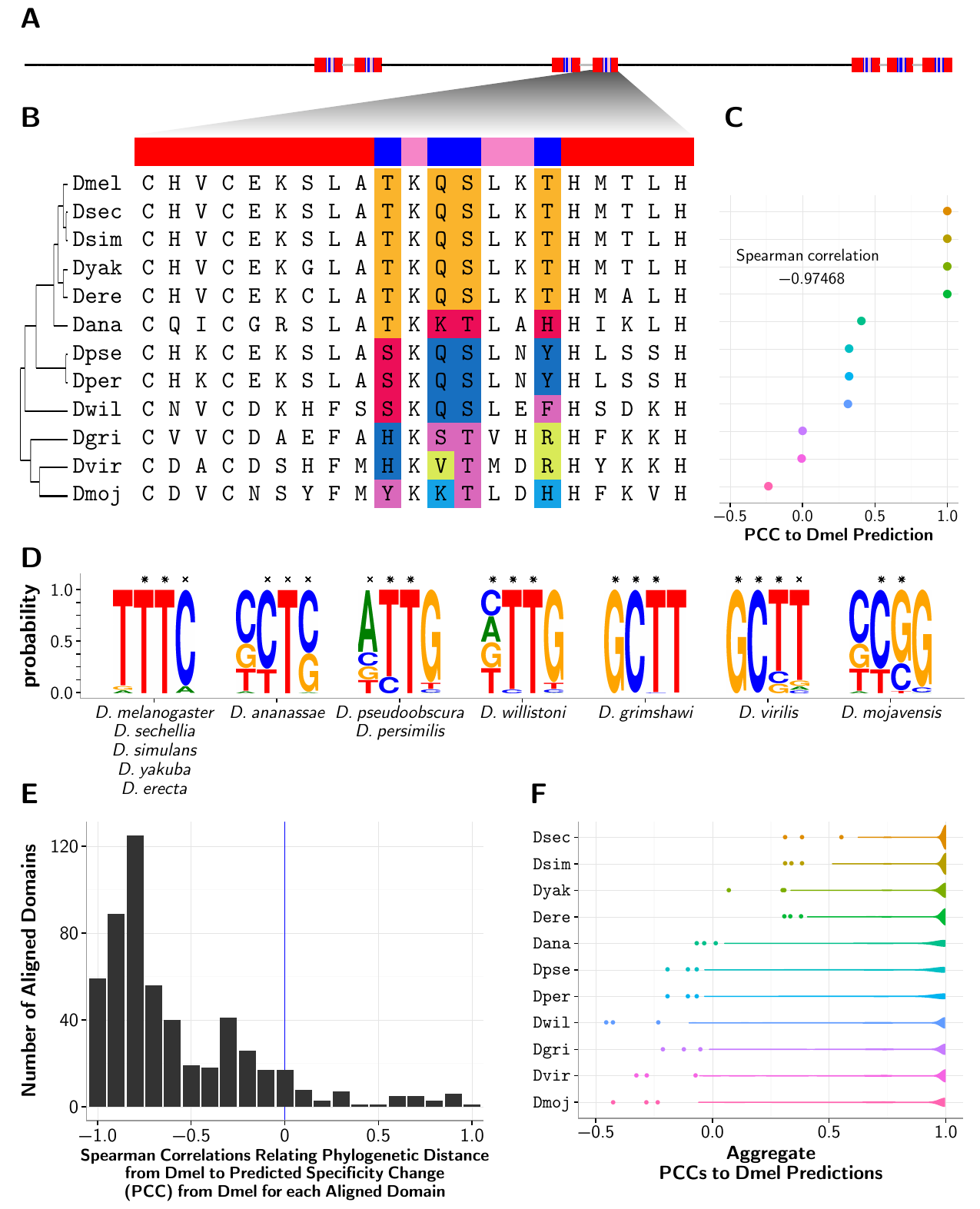}
  \end{center}
\end{adjustwidth}

\vspace{1cm}
\noindent
{\bf Figure 5. Example of a varying \emph{D. melanogaster} C2H2-ZF domain.}
  \\
       (A) Layout of the seven C2H2-ZF domains in \emph{D. melanogaster} protein FBpp0072605. All domains are found in three canonically linked arrays of sizes 2, 2, and 3 respectively. Both domains in the middle array and domains 2 and 5 located at the end of the first array and start of the last array also exhibit divergent binding residues. (B) Closeup of the 4th domain in the protein, with phylogenetic tree and multiple alignment of the aligned domains from the other fly species. (C) Average (across positions b1-b4) Pearson correlation coefficients (PCCs) between non-reference and \emph{D. melanogaster} SVM predicted specificities by species. The Spearman correlation, relating non-\emph{melanogaster} predicted specificity change to phylogenetic distance from reference \emph{D. melanogaster}, is also shown and implies that specificity changes increase gradually with distance from the reference. (D) Frequency plots of the PWMs generated by WebLogo \cite{weblogo} representing unique binding specificities, predicted by the SVM method, ordered by phylogenetic distance from \emph{D. melanogaster}, and labeled with the species whose domains had that corresponding binding specificity. Predicted positions with a PCC $>$ 0.25 to one of either the ML or RF corresponding predictions are marked with a {\sf{$\times$}}, and positions with a PCC $>$ 0.25 to both the ML and RF corresponding predictions are marked with a {\sf{$\ast$}}. (E) Distribution of Spearman correlations for each aligned domain (as in Part C) relating non-\emph{melanogaster} predicted specificity change to phylogenetic distance from reference \emph{D. melanogaster}. (F) Violin plots depicting the distributions of PCCs between predicted specificities for non-reference domains and their aligned domains in \emph{D. melanogaster} orthologs.

\newpage

\section*{Supplemental Figures and Tables}

\begin{adjustwidth}{-1in}{-1in}
  \begin{center}
    \includegraphics{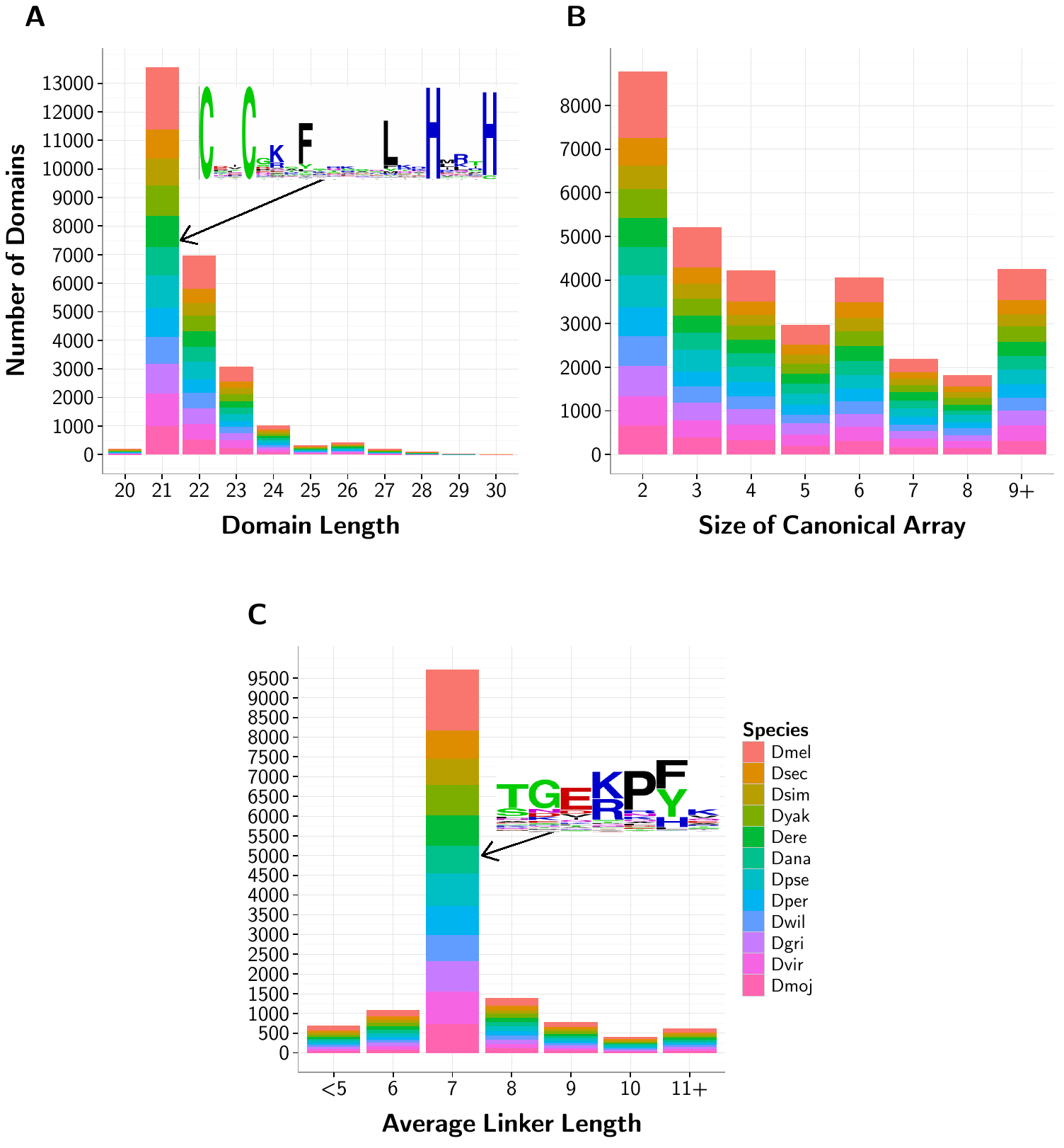}
  \end{center}
\end{adjustwidth}

\vspace{1cm}
\noindent
  {\bf Figure S1. Overview of \emph{Drosophila} C2H2-ZFs.}
  \\
    (A) Distribution of the lengths of all identified C2H2-ZF domains across all species with a sequence logo of the most common 21aa domain shown. (B) The distribution of number of domains per \emph{array}; a single protein sequence may contain multiple arrays of domains. An array is defined as adjacent C2H2-ZF domains separated by up to 12 amino acids. (C) Distribution of linker region (i.e. amino acid regions between adjacent C2H2-ZF domains) lengths with a sequence logo of the most common 7aa linker shown.

\newpage

\begin{adjustwidth}{-1in}{-1in}
  \begin{center}
    \includegraphics{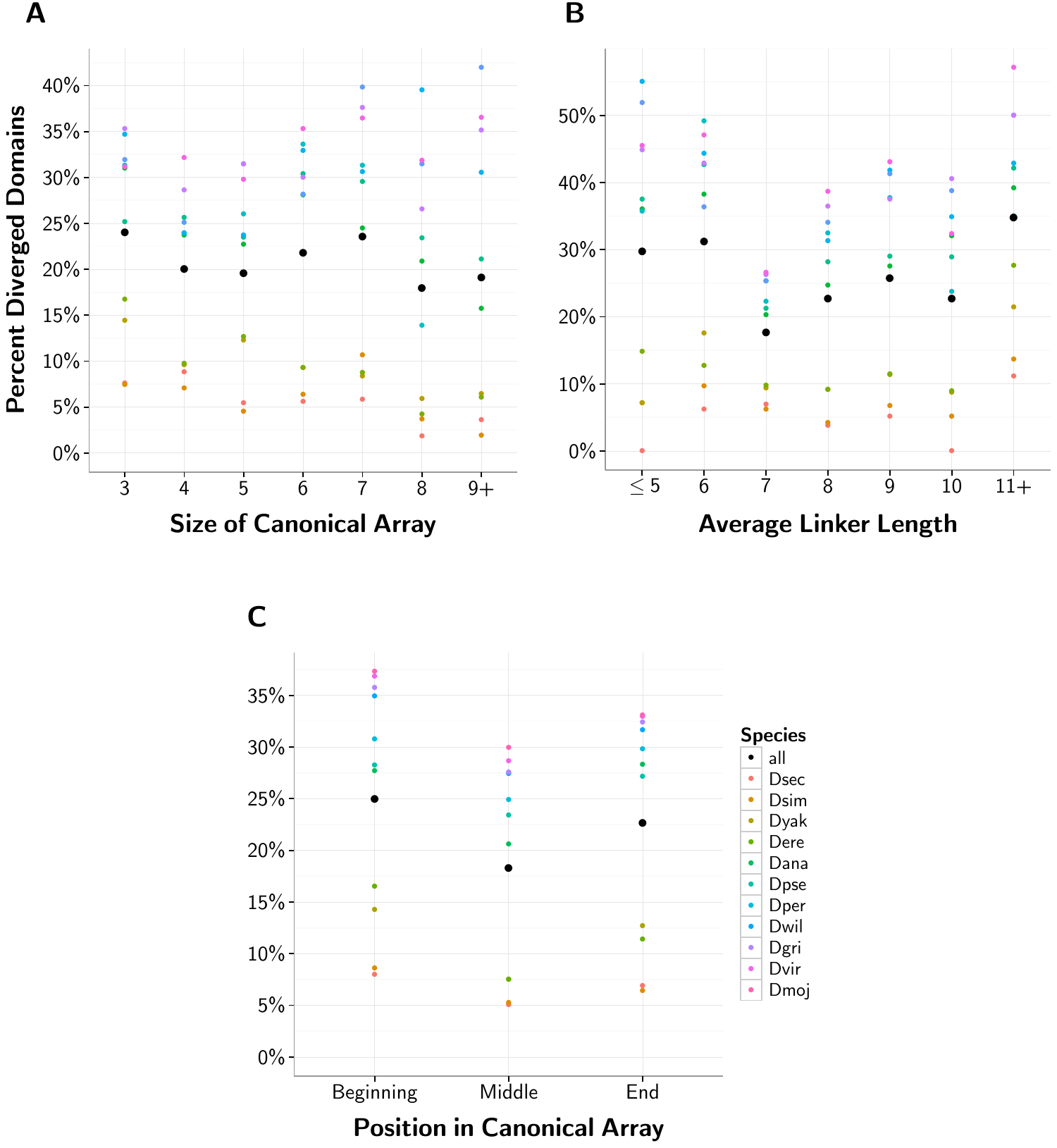}
  \end{center}
\end{adjustwidth}

\vspace{1cm}
\noindent
  {\bf Figure S2. Conservation of canonically linked C2H2-ZFs.}
  \\
    Overall and by-species divergence of aligned, canonically linked domains.  A domain is considered diverged if it differs from its corresponding aligned \emph{D. melanogaster} domain in one or more of the four specificity-determining positions -1, 2, 3, or 6.  Divergence is shown according to (A) size of tandem array in which the domain appears, (B) average length of the linker(s) bordering the domain, and (C) position (beginning, middle, or end) of the domain. A domain may only fall into one of the three categories; paired domains are labeled as beginning and end with no middle. 

\newpage

\begin{adjustwidth}{-1in}{-1in}
  \begin{center}
    \includegraphics{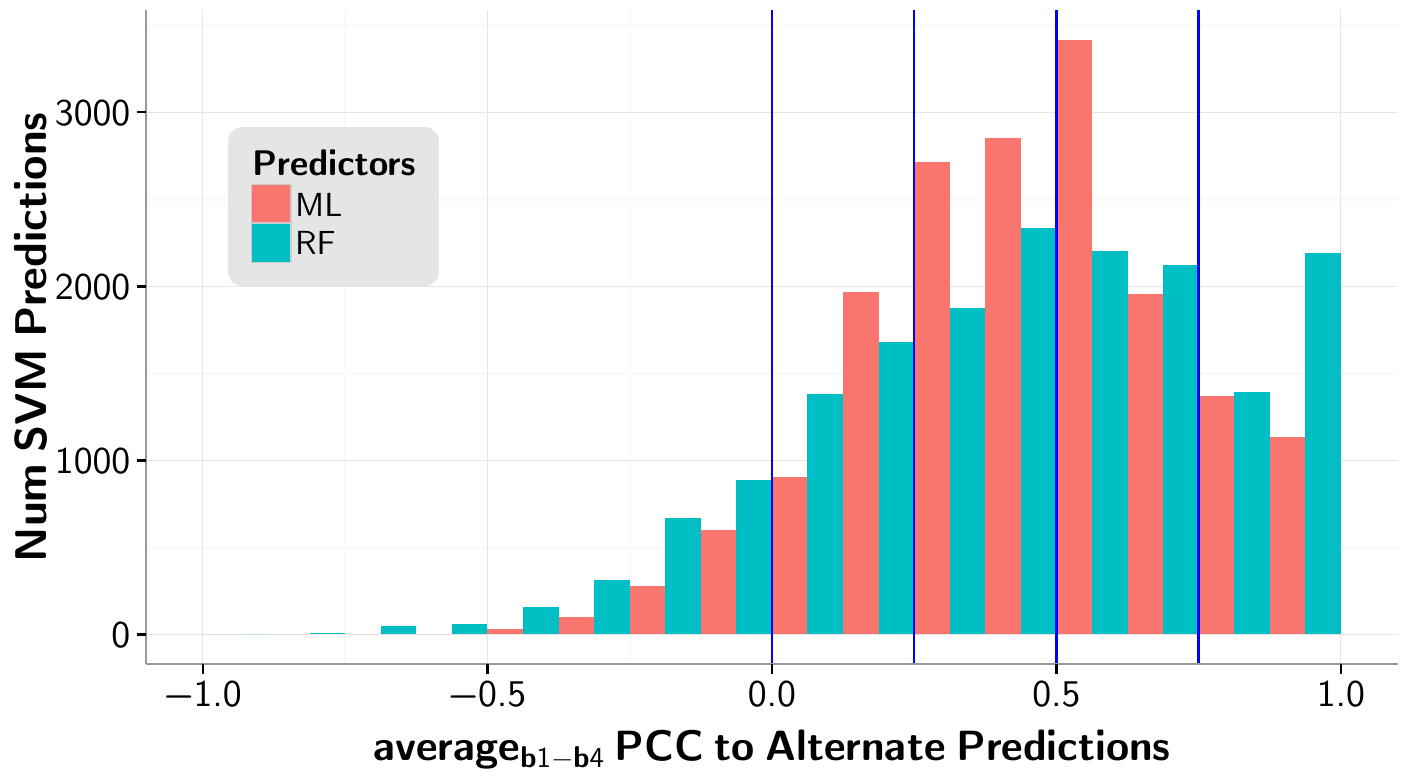}
  \end{center}
\end{adjustwidth}

\vspace{1cm}
\noindent
  {\bf Figure S3. PCCs between SVM and ML/RF predicted specificities.}
  \\
    Distribution of the PCCs between the SVM predicted binding specificity (PWM) and the ML predicted (red) and RF predicted (blue) binding specificities for all aligned domains from all \emph{Drosophila} fly species. Blue vertical lines at 0, 0.25, 0.50, and 0.75 show the thresholds used for selecting confident predictions.

\newpage

\begin{adjustwidth}{-1in}{-1in}
  \begin{center}
    \includegraphics{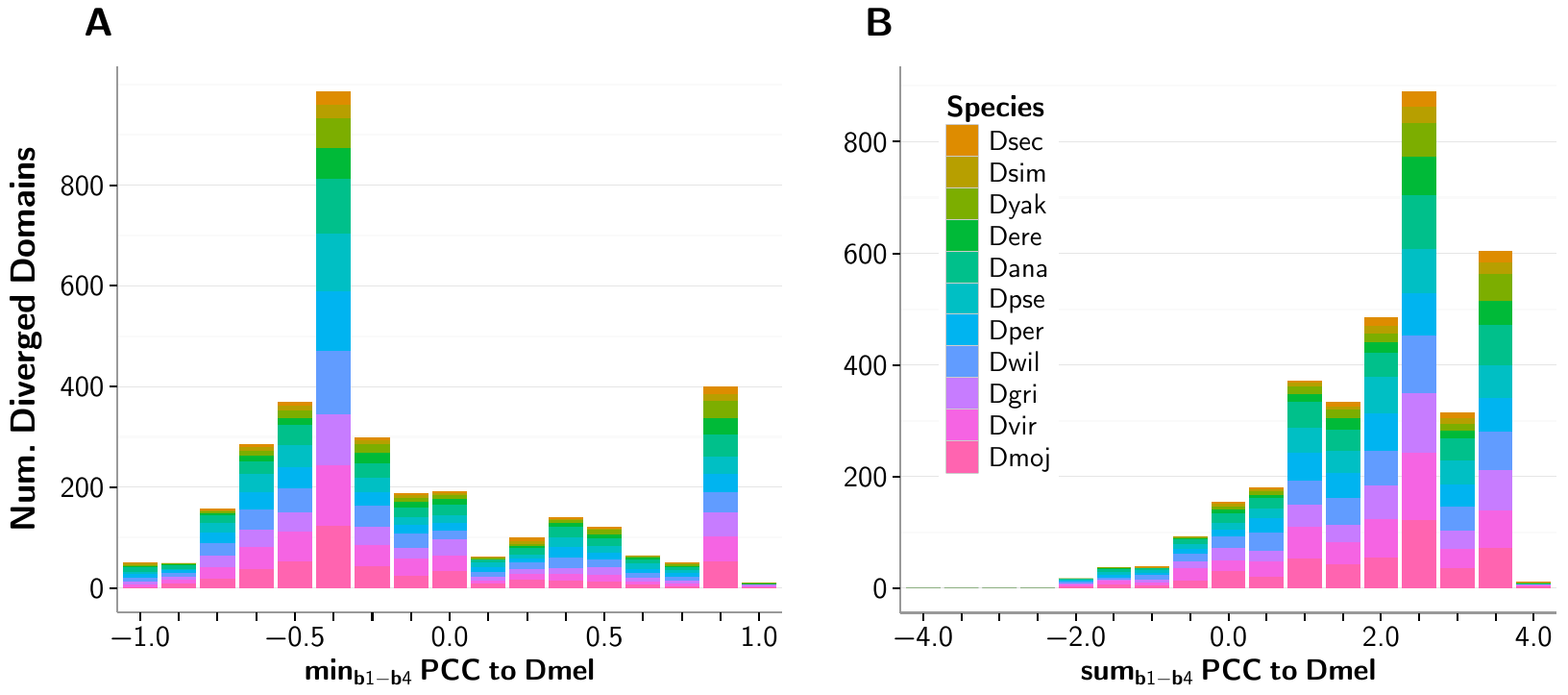}
  \end{center}
\end{adjustwidth}

\vspace{1cm}
\noindent
      {\bf Figure S4. PCCs between \emph{D. melanogaster} and non-reference predicted specificities.}
      \\
    For each divergent non-\emph{melanogaster} domain, we compared its SVM predicted specificity to the predicted specificity of its orthologous aligned \emph{D. melanogaster} domain by calculating a PCC at each position b1 through b4. (A) Distribution of divergent domains per species by minimum PCC at any one position b1 through b4 from the aligned \emph{D. melanogaster} domain. This shows that most divergent domains had a corresponding divergent binding specificity from \emph{D. melanogaster} in at least one predicted position. (B) Distribution of divergent domains per species by sum of PCCs across positions b1 through b4 from the aligned \emph{D. melanogaster} domain. All domains with a sum of PCCs $<$ 2.0 must have had a divergent binding specificity in more than one predicted position from \emph{D. melanogaster}. 

\newpage

\ 
\vspace{-1.8cm}
\begin{adjustwidth}{-1in}{-1in}
  \begin{center}
    \includegraphics{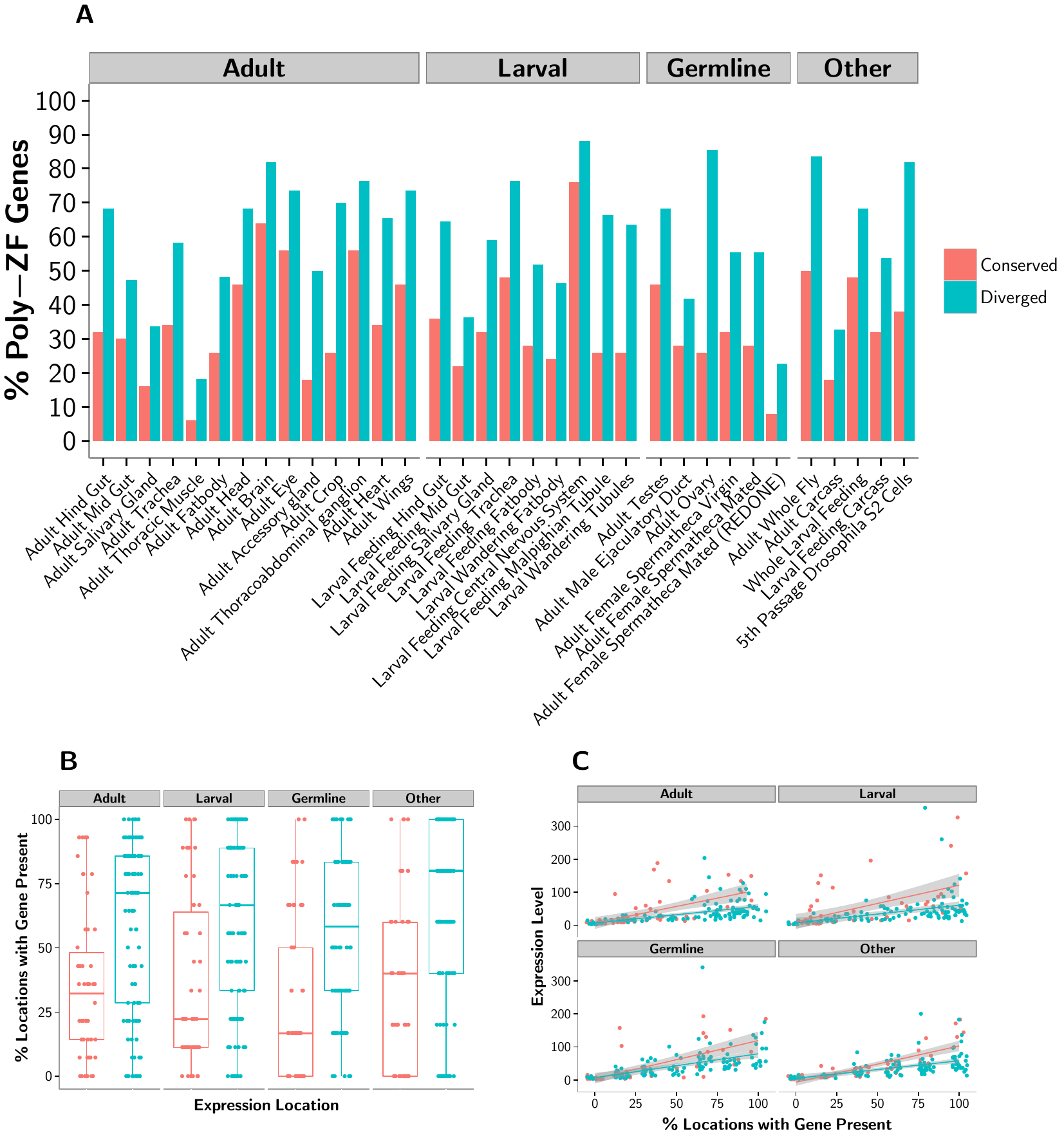}
  \end{center}
\end{adjustwidth}

\vspace{1cm}
\noindent
   {\bf Figure S5. Expression of conserved and diverged poly-ZFs by tissue.}
   \\
    (A) Percent of conserved (red) and diverged (blue) poly-ZF genes present in each tissue. FlyAtlas reports each gene as present or absent in each tissue separately across four replicates based on raw expression values \cite{flyatlas}; we consider a gene to be present in a tissue if it was marked as present across all four of these replicates. Genes were marked as present or absent in each tissue (B) Ubiquity of conserved and diverged poly-ZF genes according to the number of distinct tissues within the groups adult, larval, germline, and other they are present (binary score) in. (C) Raw expression level of each conserved and diverged poly-ZF gene by tissue type as a function of ubiquity as described in part B, with regression lines overlaid.

\newpage

\ 
\renewcommand{\tabcolsep}{0.11cm}
\vspace{-2.1cm}
\begin{center}
\rotatebox{90}{
{\sf{
\begin{tabular}{l|cccc|c||cccc|c||cccc|c} 
    \multirow{3}{*}{\textbf{}} & \multicolumn{5}{c||}{\textbf{Conservation Across Pairwise Clades}} & \multicolumn{5}{c||}{\textbf{Evolutionary Rate}} & \multicolumn{5}{c}{\textbf{Rapid Fixation}} \tn \cline{2-16}
    & \multicolumn{4}{c|}{Num. Diverged Residues} & \multirow{2}{*}{$p$-value} & \multicolumn{4}{c|}{Num. Diverged Residues} & \multirow{2}{*}{$p$-value} & \multicolumn{4}{c|}{Num. Diverged Residues} & \multirow{2}{*}{$p$-value} \tn \cline{2-5} \cline{7-10} \cline{12-15}
& -1,2,3,6 & BG & C2H2 & linker & & -1,2,3,6 & BG & C2H2 & linker & & -1,2,3,6 & BG & C2H2 & linker & \tn \hline
{\colorbox{Goldenrod}{Dsec}} & 
82 & 1288 & 157 & 55 & \emph{3.081e-01} & 
57 & 726 & 103 & 37 & \emph{1.116e-02} & 
828 & 35483 & 2787 & 1055 & 2.154e-19 \tn
{\colorbox{Goldenrod}{Dsim}} & 
80 & 1650 & 164 & 59 & \emph{5.884e-02} & 
44 & 706 & 63 & 36 & 7.452e-04 & 
860 & 31833 & 2876 & 1029 & 1.077e-16 \tn
{\colorbox{YellowGreen}{Dyak}} & 
244 & 7264 & 722 & 268 & \emph{2.705e-01} & 
122 & 3011 & 305 & 113 & 2.748e-09 & 
1893 & 85592 & 6576 & 2117 & 2.207e-23 \tn
{\colorbox{YellowGreen}{Dere}} & 
238 & 7124 & 706 & 255 & \emph{8.818e-02} & 
124 & 3335 & 302 & 95 & 1.547e-13 & 
1851 & 89557 & 6813 & 2170 & 4.195e-31 \tn
Dana & \multicolumn{4}{c|}{---} & & 
352 & 14926 & 1197 & 363 & 1.541e-39 & 
5413 & 223373 & 18980 & 6047 & 3.108e-114 \tn
{\colorbox{Peach}{Dpse}} & 
728 & 26594 & 2459 & 828 & 9.951e-04 & 
372 & 18980 & 1327 & 439 & 1.196e-43 & 
7245 & 264741 & 25376 & 8443 & 1.248e-93 \tn
{\colorbox{Peach}{Dper}} & 
730 & 27139 & 2468 & 836 & 2.051e-07 & 
364 & 17195 & 1286 & 392 & 2.313e-35 & 
4838 & 183933 & 18820 & 6233 & 6.566e-37 \tn
Dwil & \multicolumn{4}{c|}{---} & & 
499 & 22041 & 1600 & 489 & 8.597e-56 & 
6357 & 258351 & 23266 & 7270 & 1.827e-125 \tn
Dgri & \multicolumn{4}{c|}{---} & & 
453 & 21009 & 1600 & 518 & 6.579e-46 & 
4884 & 226709 & 19313 & 6302 & 3.676e-52 \tn
{\colorbox{CornflowerBlue}{Dvir}} & 
852 & 30927 & 2853 & 956 & 1.604e-22 & 
469 & 21682 & 1530 & 516 & 1.274e-51 & 
6952 & 256468 & 24503 & 8393 & 3.469e-78 \tn
{\colorbox{CornflowerBlue}{Dmoj}} & 
887 & 31491 & 3008 & 981 & 1.139e-19 & 
476 & 23032 & 1647 & 514 & 9.714e-49 & 
5980 & 222726 & 22225 & 7582 & 3.772e-54 \tn
    \multicolumn{16}{l}{} \tn 
    \multicolumn{16}{l}{} \tn 
    \multicolumn{16}{l}{\rm\bf Table S1. Divergent Residue Counts and Significance Values} \tn
    \multicolumn{16}{l}{\rm Counts of divergent residues and significance between binding residues and background per non-reference species used for the three} \tn
    \multicolumn{16}{l}{\rm calculations of functional importance previously described: (Major Column 1) Conservation Across Clades, (Major Column 2) Evolutionary} \tn
    \multicolumn{16}{l}{\rm  Rate, and (Major Column 3) Rapid Fixation. The first four subcolumns within each major column correspond to the number of divergent} \tn
    \multicolumn{16}{l}{\rm residues in specificity-conferring positions -1, 2, 3, and 6 in C2H2-ZF domains ({\sf{-1,2,3,6}}), background divergent residues outside of arrays of} \tn
    \multicolumn{16}{l}{\rm canonically linked domains ({\sf{BG}}), residues outside of the alpha-helix within C2H2-ZF domains ({\sf{C2H2}}), and linker regions between adjacent} \tn
    \multicolumn{16}{l}{\rm canonically linked domains ({\sf{linker}}). The fifth subcolumn in each major column is the exact $p$-value comparing the {\sf{-1,2,3,6}} residues to {\sf{BG}}} \tn
    \multicolumn{16}{l}{\rm residues using a binomial test in major columns 1 and 3 and wilcoxon test in major column 2. In Major Column 1, residues are only included} \tn
    \multicolumn{16}{l}{\rm from each species where there is a non-gapped residue of the same type (e.g. {\sf{-1,2,3,6}}, {\sf{BG}}, {\sf{C2H2}}, {\sf{linker}}) in an ortholog of its partner species.} \tn
    \multicolumn{16}{l}{\rm In Major Column 2, only complete (i.e. an ortholog from all 12 fly species) columns are included, as measurements of evolutionary rate using} \tn
    \multicolumn{16}{l}{\rm less complete multiple alignments could not be compared. In Major Column 3, all divergent residues, regardless of alignment to any species} \tn
    \multicolumn{16}{l}{\rm apart from the reference \emph{D. melanogaster} are included.} \tn
    \multicolumn{16}{l}{\rm } \tn
\end{tabular} 
}}}
\end{center}

  \newpage
  
\renewcommand{\tabcolsep}{0.2cm}  

  \begin{adjustwidth}{-1in}{-1in}
   \begin{center}
  {\sf{
  \begin{tabular}{>{\centering\arraybackslash}m{1.65in} | 
                   >{\centering\arraybackslash}m{0.72in} | 
                   >{\centering\arraybackslash}m{0.79in} | 
                   >{\centering\arraybackslash}m{0.79in} | 
                   >{\centering\arraybackslash}m{0.79in} |                    
                   >{\centering\arraybackslash}m{0.72in}}
  & \multicolumn{5}{c}{\textbf{Confidence Thresholds}} \tn \cline{2-6}
  & \small{\textbf{None}}
  & \small{\textbf{PCC $\ge$ 0}}
  & \small{\textbf{PCC $\ge$ 0.25}}
  & \small{\textbf{PCC $\ge$ 0.5}}
  & \small{\textbf{PCC $\ge$ 0.75}} \tn \hline
\textbf{Total Domains} & 17319 & 16893 (97.5\%) & 15098 (87.2\%) & 10742 (62.0\%) & 4978 (28.7\%) \tn \hline
\textbf{Total Dmel Domains with 1{\tt{+}} Binding Residue Changes in Other Flies} & 797 & 768 (96.4\%) & 655 (82.2\%) & 389 (48.8\%) & 112 (14.1\%) \tn \hline
\textbf{Total Aligned Domains with Spearman $<$ 0} & 714 (89.6\%) & 689 (89.7\%) & 582 (88.9\%) & 347 (89.2\%) & 96 (85.7\%) \tn \hline
\textbf{Total Aligned Domains with Spearman $<$ -0.5} & 517 (64.9\%) & 499 (65.0\%) & 428 (65.3\%) & 264 (67.9\%) & 79 (70.5\%) \tn \blank{\hline
\textbf{Gradient of Linear Regression Line Relating Specificity Change (PCC) to LCA from Dmel} & -0.02256 & -0.02153 & -0.01792 & -0.01335 & -0.00656 \tn \hline}
\end{tabular}}}
   \end{center}
  \end{adjustwidth}

\vspace{1cm}
\noindent
 {\bf Table S2. Binding specificity change (PCC) from reference for various confidence thresholds.} 
 \\ 
Change in binding specificity for aligned domains between the reference \emph{D. melanogaster} and non-reference fly species. Confidence is measured by comparing SVM predicted specificities to those produced by ML and RF methods using PCC. The domains included at each confidence threshold are those where the SVM predicted specificities were within a particular PCC cutoff when compared to either the ML or RF predicted specificities. (Row 1) Total number of aligned, ungapped domains across all 12 fly species. (Row 2) Total number of ungapped \emph{D. melanogaster} domains that have 1-to-1 orthologs in at least 1 other fly and exhibit a divergent binding residue in at least 1 other fly. (Row 3) Total number of aligned orthologous domains across \emph{D. melanogaster} and at least 1 other fly species where the Spearman correlations relating the phylogenetic distance to \emph{D. melanogaster} for each non-reference fly domain to the change in predicted specificity from the aligned \emph{D. melanogaster} domain (measured using PCC) is less than 0. (Row 4) Total number of aligned orthologous domains where the Spearman correlations, as in Row 3, are less than -0.5. 

\vspace{3cm}
\noindent
{\tt http://compbio.cs.princeton.edu/zfvariation/tables3.txt}

\vspace{0.5cm}
\noindent
{\bf Table S3. Enrichment of Gene Ontology terms by GO TermFinder.}
\\
GO Term Enrichment for \emph{Drosophila} conserved and diverged poly-ZF genes as generated by GO Term Finder (go.princeton.edu).

\end{document}